\documentclass[letterpaper,twocolumn,10pt]{article}
\usepackage{usenix2019,epsfig,endnotes}
\usepackage{amsthm}
\usepackage{amsfonts}
\usepackage{amsmath}
\usepackage{xcolor}
\usepackage{textcomp}

\usepackage{bm}
\usepackage{mathrsfs}

\usepackage{enumitem}
\usepackage{listings}
\usepackage{threeparttable}
\usepackage{multirow}
\usepackage{flushend}
\usepackage{graphicx}
\usepackage{url}
\usepackage{authblk}
\newcommand{\tabincell}[2]{\begin{tabular}{@{}#1@{}}#2\end{tabular}}
\newtheorem{theorem}{Theorem}

\usepackage{amssymb}
\usepackage{pifont}
\newcommand{\cmark}{\ding{51}}%
\newcommand{\xmark}{\ding{53}}%
\begin{document}

\date{}

\title{\Large \bf FALCON: A Fourier Transform Based Approach for Fast and Secure Convolutional Neural Network Predictions}

\author[1]{Shaohua Li}
\author[1]{Kaiping Xue}
\author[1]{Chenkai Ding}
\author[1]{Xindi Gao}
\author[2]{David S L Wei}
\author[3]{Tao Wan}
\author[1]{Feng Wu}
\affil[1]{Department of EEIS, University of Science and Technology of China, Hefei, Anhui 230027, China}
\affil[2]{Department of Computer and Information Science, Fordham University, New York, NY 10458, USA}
\affil[3]{CableLabs, 858 Coal Creek Circle, Louisville, Colorado, USA }

\maketitle


\subsection*{Abstract}
Machine learning as a service has been widely deployed to utilize deep neural network models to provide prediction services. However, this raises privacy concerns since clients need to send sensitive information to servers. In this paper, we focus on the scenario where clients want to classify private images with a convolutional neural network model hosted in the server, while both parties keep their data private. We present FALCON, a fast and secure approach for CNN predictions based on Fourier Transform. Our solution enables linear layers of a CNN model to be evaluated simply and efficiently with fully homomorphic encryption. We also introduce the first efficient and privacy-preserving protocol for softmax function, which is an indispensable component in CNNs and has not yet been evaluated in previous works due to its high complexity. We implemented the FALCON and evaluated the performance on real-world CNN models. The experimental results show that FALCON outperforms the best known works in both computation and communication cost.

\section{Introduction}
Deep learning has been applied to quite a few fields to overcome the limitations of traditional data processing methods, such as image classification \cite{krizhevsky2012imagenet, he2016deep}, speech recognition \cite{abdel2014convolutional, amodei2016deep}, medical diagnosis \cite{shen2017deep, esteva2017dermatologist}, etc. Some companies and institutions have also invested in deep learning technologies, and trained their own deep neural network models to provide users with paid or free services. For example, Google Vision \cite{googlevision} provides an API for image classification for developers and general users, and one can upload an image to the cloud to obtain the classification result and its corresponding probability.
Although these services provide rich experiences to users, they also cause serious privacy concerns because uploaded user data may contain private information \cite{shokri2015privacy}, such as face pictures and X-ray images. Although many companies claim that they will never leak or use users' data for commercial purposes, the increasing number of user data leaks alert us that there is no guarantee on what they promised~\cite{abadi2016deep}.

Certainly, the clients' input data is not the only sensitive information, because for servers, their own models also need to be protected from adversarial clients.
First, models may be trained with large amount of private data, e.g., medical records to obtain a model for disease prediction. Thus, sensitive information could be extracted from a trained model if disclosed to a malicious client \cite{shokri2017membership, song2017machine}. Second, model parameters and detailed prediction results, i.e., accurate probabilities over all classes, can be used to generate adversarial examples to deceive deep learning models \cite{sharif2016accessorize, carlini2017towards}, to result in incorrect classification results.
Third, many prediction models themselves, even without considering sensitive training data, require intellectual property protection and cannot be disclosed to third parties including their clients \cite{liu2017oblivious, juvekar2018gazelle}. With the continuous development of deep learning technology, the privacy protection for user data and models has become a key issue needed to be thoroughly addressed in  machine learning as a service.

To tackle this problem, researchers have put forward a secure deep learning scenario, where the server has a model, the client has data, and these two interact in such a way that the client can obtain the prediction result without leaking anything to the server, while learning nothing about the model. When we regard a deep neural network model as a function, an intuitive idea for protecting privacy is to use homomorphic encryption to evaluate it. A typical representative is the CryptoNets \cite{gilad2016cryptonets} proposed by Gilad-Bachrach et al., who used leveled homomorphic encryption (LHE) \cite{bos2013improved} to evaluate the entire neural network in the ciphertext domain. The limitations of CryptoNets are its heavy computation overhead and required modification of neural networks' structure. Secure multi-party computing is another path for secure function evaluation, and one of the representatives is MiniONN \cite{liu2017oblivious} proposed by Liu et al., which transforms a neural network model into an oblivious form and evaluates it with secure two-party computation. Most recently, Juvekar et al. proposed GAZELLE \cite{juvekar2018gazelle}, which used homomorphic encryption and secure two-party computing to obtain better performance.

In this paper, we focus on the fast and secure solution for Convolutional Neural Networks (CNNs), one of the most important neural networks in deep learning. This kind of neural network is characterized by the spatial input data, such as images and speeches. Typically, a deep neural network consists of many layers, each of which has its specific functionality. For a CNN model, it usually includes convolutional, activation, pooling, and fully-connected layers, and often follows by a softmax layer. Convolutional and fully-connected layers have linear property, while activation and pooling are nonlinear layers. The softmax layer is used to normalize the output of previous layers into the form of probability representation, usually used in the last layer of a CNN to output a human-friendly prediction result. The softmax layer is indispensable in many use cases. For example, a CNN model classifies an X-Ray image into ``Pneumonia'' with probability 5\%. Although ``Pneumonia'' is the top label, the real result actually indicates that the patient has no such disease, and this cannot be known without the probability output. Because the softmax function involves division and exponentiation that are very difficult to evaluate in a privacy-preserving way, the existing works, e.g., CryptoNets, MiniONN and GAZELLE, used argmax function instead of softmax to obtain only the top one label. We, however, propose a novel efficient protocol for the softmax layer.

In FALCON, we utilize the linear property of fast Fourier Transform (FFT) to design simple but efficient protocols for linear layers, i.e., convolutional and fully-connected layers. Our design enables the fully homomorphic encryption to be used in linear layers in a straightforward way, thereby significantly improving the evaluation performance.
For non-linear activation and pooling layers, we use Yao's Garbled Circuits \cite{yao1986generate} to implement secure computation. We provide detailed evaluation methods for ReLU activation function and max pooling function in FALCON. Those functions are chosen for their popularity in CNN. For the privacy-preserving evaluation of softmax layer, which has not yet been addressed by any previous work, we present an efficient and secure two-party computation protocol, which is free of using expensive division and exponentiation computations in boolean circuits.

Our contributions are summarized as follows:
\begin{itemize}[leftmargin=17pt]
	\item We develop fast and secure protocols for convolutional and fully connected layers. Our fast Fourier Transform based design keeps the utilization of fully homomorphic encryption simple but effective. We present FALCON with the proposed protocols to realize an efficient and privacy-preserving evaluation for convolutional neural networks.
	\item We propose the first efficient secure two-party protocol for the softmax function with high accuracy and low computation complexity. The protocol can be used not only in FALCON, but also in other neural networks with the same security model.
	\item We implement FALCON, evaluate its performance, and compare with prior works. Our results show that the proposed scheme outperforms the best existing solutions on real-word models.
\end{itemize}

\section{Related Work}
CryptoNets \cite{gilad2016cryptonets} scheme inspired us to process neural network models securely with leveled homomorphic encryption (LHE). It is non-interactive and allows a client to obtain a result in one round of communication. Since only LHE is used, CryptoNets only supports linear activation and pooling functions, i.e. square activation and mean pooling. However, in real-world scenarios, the square activation functions are barely used for training or prediction.
SecureML \cite{mohassel2017secureml} proposed efficient and privacy-preserving protocols for training several kinds of machine learning models, including linear regression, logistic regression and neural network, between two non-colluding servers. It mainly used secure two-party computation to train various models, and Paillier cryptosystem was used to accelerate some calculations. SecureML is the first to realize privacy preserving training on neural networks, which supports predictions naturally.

DeepSecure \cite{rouhani2017deepsecure} used Yao's Garbled Circuits only to enable scalable execution of deep neural network models in a privacy-preserving setting. It focused on the scenario where semi-trusted client and server interact with each other to evaluate a trained model. Chameleon \cite{riazi2018chameleon}  comprehensively used additively secret sharing \cite{bogdanov2008sharemind}, Yao's Garbled Circuits \cite{yao1986generate}, and GMW protocol \cite{goldreich1987play} to implement efficient secure function evaluation, including CNN models.

MiniONN \cite{liu2017oblivious} transformed a neural network model into an oblivious version, and used lattice-based additively homomorphic encryption to generate multiplication triplets first, and then evaluated the model using secure two-party computation efficiently.
The more recent work GAZELLE \cite{juvekar2018gazelle} utilized the lattice-based homomorphic encryption and designed efficient schemes for privacy-preserving convolution and matrix-vector multiplication operations. GAZELLE used homomorphic cryptography in linear convolutional and fully-connected layers, and secure two-party computation in non-linear activation and pooling layers. Because MiniONN and GAZELLE outperform all previous works, we only compare FALCON with them to show our performance superiority.

In all previous works, only the SecureML mentioned that the softmax function can be evaluated with secure two-party computation. But it provided no details and we will show that the softmax max is too complex to implement with secure two-party computation directly. Other works used a compromise way to deal with it, i.e., using argmax to replace softmax function. However, often there are cases that clients need to learn the probabilities of different output labels.

\section{Preliminaries}

\begin{figure*}
	\centering
	\includegraphics[width=0.98\textwidth]{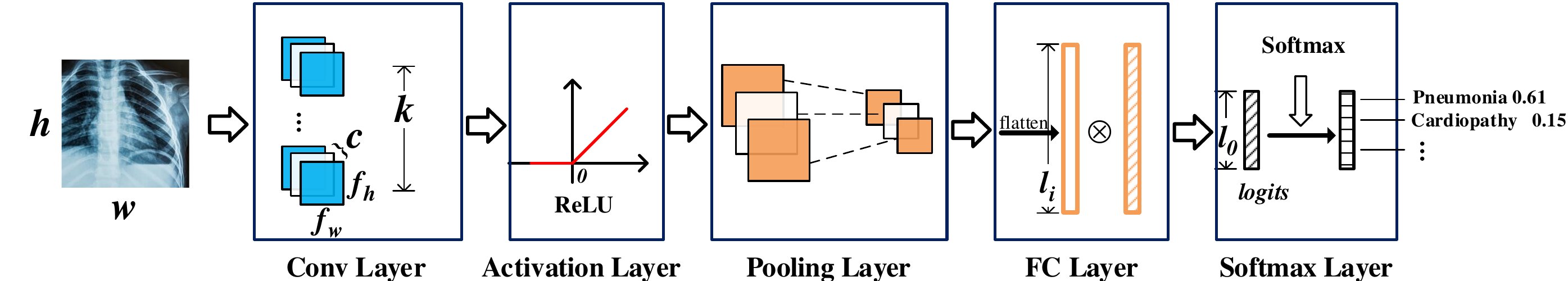}
	\caption{An example of convolutional neural networks.}
	\label{fig-CNN}
\end{figure*}

\subsection{Convolutional Neural Network}
A convolutional neural network (CNN) consists of four types of layers, namely convolutional, activation, pooling, and fully connected. An example CNN is shown in Fig. \ref{fig-CNN}. The first layer of a neural network takes the original data (e.g. an X-Ray image) as the input, and each subsequent layer computes a function over the previous output. A CNN processes the input with a sequence of layers and outputs a result vector (called logits), which demonstrates the relative weights of different potential classes. The class with the largest weight is the target class. In most situations, we will use $softmax$ function to convert the logits to a probability vector, each element of which shows the probability that the input is classified into a certain class. Below we describe the clean abstractions of each layer.

\begin{figure}
	\centering\includegraphics[width=0.48\textwidth]{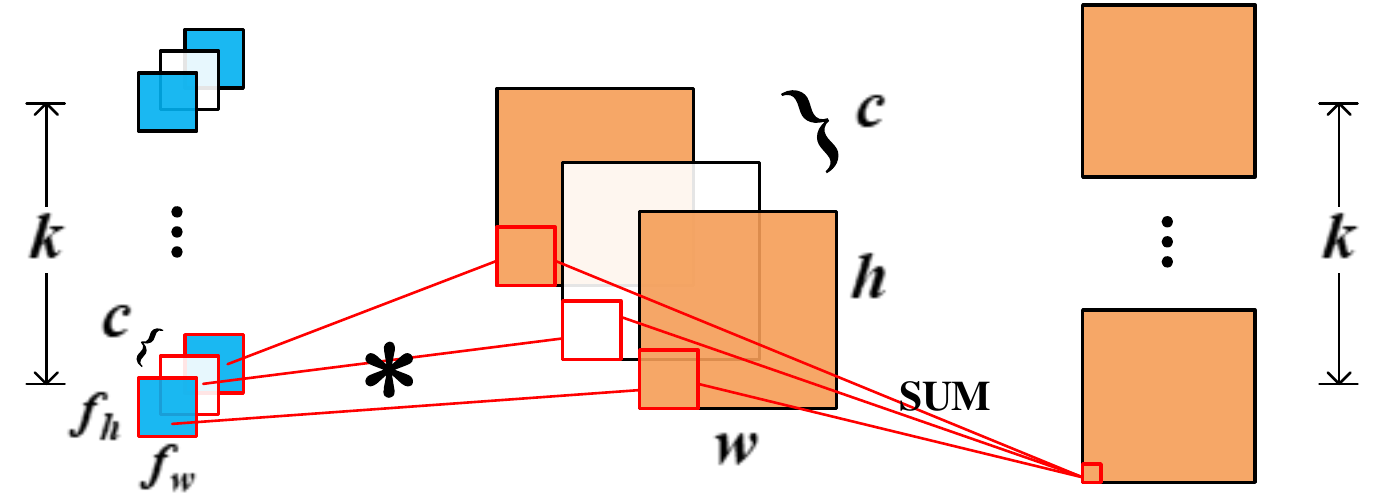}
	\caption{Convolution operation.}
	\label{fig-conv}
\end{figure}

\smallskip
\noindent\textbf{Convolutional Layer (Conv). }
The input to a convolutional layer is a $w \times h \times c$ image where $w$ and $h$ are the width and height of the image and $c$ is the number of channels, e.g. an RGB image has $c=3$. The convolutional layer has $k$ filters (or kernels) of size $f_w \times f_h \times c$. In a convolutional layer, an input image is convolved in turn with $k$ filters, and thus producing $k$ output images. To better understand this operation, we consider a $f_w \times f_h \times c$ filter convolves with a $w \times h \times c$ image. As shown in Fig. \ref{fig-conv}, for each one of $c$ images of size $w \times h$, a corresponding filter among $c$ filters of size $f_w \times f_h$ convolves with it, producing $c$ intermediate images. The final output is the result of summing up these images. Denoting the input images as $x_i,i\in [1,c]$, the filters as $w_{i,j},i\in [1,c],j\in [1,k]$, and the output images as $y_j, j\in[1,k]$, we can express the above computation as follows:
\begin{equation}
y_j=\sum_{i=1}^{c}{x_i * w_{i,j}},
\end{equation}
where ``$*$'' is convolution operation.

\smallskip
\noindent\textbf{Activation Layer. }
Neural networks use non-linear activation functions to improve non-linearity of models. Since ReLU (rectified linear unit) \cite{nair2010rectified} is the default activation function and has the best performance in popular CNNs \cite{hadji2018we}, we use it in our FALCON. The ReLU function is shown as follows:
\begin{equation}\label{relu}
f(x)=\mbox{max}\left(0,x\right).
\end{equation}
The output of a ReLU activation layer has the same shape as the input, which means $f(x)$ is applied to each pixel in the input images.

\smallskip
\noindent\textbf{Pooling Layer. }
Pooling functions reduce the dimensionality of the input by calculating the mean or the maximum value. A pooling layer will partition the input into a set of non-overlapping rectangle sub-areas and perform the mean or the maximum function on each sub-area. These two types of functions correspond to mean pooling and max pooling, respectively.

\smallskip
\noindent\textbf{Fully Connected Layer (FC). }
The input to a fully connected layer is a vector $\bm{v_i}$ of length $l_i$, and after multiplying with a weight matrix $\bm{W}$ of size $l_o \times l_i$, it outputs a vector $\bm{v_o}$ of length $n_o$, shown as follows:
\begin{equation}
\bm{v_o} = \bm{W}\bm{v_i}+\bm{b},
\end{equation}
where $\bm{b}$ is a bias vector of length $l_o$ and can be discarded in some networks. Note that the original input is a group of images, so we need to flatten them before feeding to a fully connected layer.

\smallskip
\noindent\textbf{Softmax Function. }
In general, the output of a CNN is called $logits$, which shows the scores of different classes for an input. In image classification scenarios, a softmax layer will be used as the last layer to convert the $logits$ to the probabilities using
\begin{equation}
f(\bm{x})_k=\frac{e^{x_i}}{\sum_{k=1}^{K}{e^{x_k}}}, \quad\mbox{for $i=1,2,\cdots,K$},
\end{equation}
where $K$ is the number of classes as well as the length of $logits$, and we can find that
$\sum_{k=1}^{K}f(\bm{x})_k=1$. In most scenarios, we only need the classification result $t$, where $t=\mbox{argmax}(logits)$, and the corresponding probability $f(\bm{x})_t$. In all previous work, such as MiniONN and GAZELLE, clients can only obtain $t$, while in our FALCON design, $f(\bm{x})_t$ is also available to clients.

\subsection{Fast Fourier Transform}
In image processing, the well-known fast Fourier Transform (FFT) is an algorithm that can convert an image from its space domain to a representation in the frequency domain and vice versa \cite{winograd1978computing}. Letting $f(x,y)$ denote the pixel value of a $M \times N$ image at point $(x,y)$, we can compute the FFT of the image as follows:
\begin{equation}\label{fft}
\mathcal{F}_f(u,v) = \sum_{x=0}^{M-1}\sum_{y=0}^{N-1}{f(x,y) \cdot e^{-j2\pi(\frac{ux}{M} + \frac{vy}{N})}},
\end{equation}
where $u\in[0, M-1], v\in[0,N-1]$, and $\mathcal{F}_f(u,v)$ is a complex number. From the formula we can see that the representations of an image in different domains have the same shape. For simplicity, we denote the FFT of input \textbf{x} as $\mathcal{F}(\textbf{x})$. An important property of the FFT used in this paper is \textit{linearity}, i.e. for two inputs \textbf{x} and \textbf{y}, we have the following equation:
\begin{equation}\label{fft_linearity}
\mathcal{F}(\textbf{x}) + \mathcal{F}(\textbf{y}) = \mathcal{F}(\textbf{x}+\textbf{y}).
\end{equation}

To reduce the number of required operations when evaluating convolutional layers, we use the \textit{Convolution Theorem}: the convolutions in the space domain are equivalent to pointwise products in the frequency domain. Denoting $\mathcal{F}^{-1}$ as the inverse FFT, the convolutions between inputs \textbf{x} and \textbf{y} can be computed as follows:
\begin{equation}\label{conv_theorem}
\textbf{x} * \textbf{y} = \mathcal{F}^{-1}(\mathcal{F}(\textbf{x}) \cdot \mathcal{F}(\textbf{y})).
\end{equation}
This transformation can be used to accelerate training and prediction of CNN models. The basic idea is to convert the input and filters of a convolutional layer to the frequency domain, and then perform the pointwise product instead of the convolution. In this paper, we use the similar idea to speed up the convolution operation in the ciphertext domain, and we will show that FFT operations have little effect on the performance of a CNN model.

\subsection{Cryptographic Building Blocks}
\subsubsection{Lattice-based Homomorphic Encryption.}
To implement privacy-preserving convolutions in FALCON, we require two kinds of homomorphic operations: SIMDAdd and SIMDMul. The SIMD here means we can pack a vector of plaintext elements into a ciphertext, and perform calculations on ciphertexts corresponding to each plaintext element, which reduces required ciphertext size and evaluation time. The SIMDAdd represents homomorphic addition between two ciphertexts, while the SIMDMul represents homomorphic multiplication between a ciphertext and a plaintext. All these requirements can be satisfied by modern lattice-based homomorphic encryption systems \cite{gentry2012fully, fan2012somewhat, brakerski2014leveled}. There are three required parameters in these schemes, namely number of plaintext slots $n$, plaintext module $p$, and ciphertext module $q$. Parameter $n$ is the maximum number of data that can be processed in SIMD style. Parameter $p$ limits the range of plaintext data. Parameter $q$ can be calculated from given $n$ and $p$. In this paper, we denote the ciphertext of $x$ as $\left[x\right]$.

\subsubsection{Secure Two-Party Computation.}
Secure two-party computation protocols allow two parties to jointly evaluate functions on each other's private data while preserving their privacy. Functions are represented as \textit{boolean circuits} and then computed by these protocols. Yao's Garbled Circuits is a representative implementation of such protocols and will be used in this paper for the secure computation between a client and a server.

The ABY framework \cite{demmler2015aby} is an open source library that supports secure two-party computation, and we use this library to implement secure ReLU, max pooling and softmax layer.
This library has encapsulated several basic operations for secure computation, and we here introduce the operations used in this paper (Note that, the term ``share' used in what follows means Yao sharing, which is a type of sharing used by Yao's Garbled Circuits.):
\begin{itemize}[leftmargin=10pt]
	\item[--]\textit{ADDGate(a, b)} performs an arithmetic addition on input shares $a$ and $b$, and returns the result as a share.
	\item[--]\textit{SUBGate(a, b)} performs an arithmetic subtraction on input shares $a$ and $b$, and returns the result as a share.
	\item[--]\textit{GTGate(a, b)} performs a ternary operation ``$a > b~?~1 : 0$'', and returns 1 if $a > b$, 0 otherwise.
	\item[--]\textit{MUXGate(a, b, s)} performs as a multiplexer, and returns $a$ if $s$ is 1, returns $b$ if $s$ is 0.
\end{itemize}
Since the ABY supports SIMD technique, all these operations can receive arrays as inputs, and process them in SIMD style, which can reduce the memory footprint and evaluation time of circuits.

\begin{table*}
	\begin{threeparttable}
		\centering
		\caption{Privacy guarantees in related works.}
		\label{table:privacy}
		\begin{tabular}{|c|c|c|c|c|c|c|c|}
			\hline
			& input content & input size & model weights & type of layer & layer size & number of layers  & filter size \\
			\hline
			\hline
			FALCON & \cmark & \xmark & \cmark & partial & \xmark & \xmark & \cmark \\
			\hline
			GAZELLE \cite{juvekar2018gazelle} & \cmark & \xmark & \cmark & partial & \xmark & \xmark & \cmark \\
			\hline
			MiniONN \cite{liu2017oblivious} & \cmark & \xmark & \cmark & \xmark & \xmark & \xmark & \xmark \\
			\hline
			SecureML \cite{mohassel2017secureml} & \cmark & \xmark & \cmark & \xmark & \xmark & \xmark & \xmark \\
			\hline
			DeepSecure \cite{rouhani2017deepsecure} & \cmark & \xmark & \cmark & \xmark & \xmark & \xmark & \xmark \\
			\hline
		\end{tabular}
		\begin{tablenotes}
			\small
			\item ``\cmark'' means the scheme hides the corresponding information.
			\item ``\xmark'' means the scheme leaks the corresponding information.
			\item ``partial'' means the scheme can hide the information on which layers are convolutional layers and which are fully connected.
		\end{tablenotes}
	\end{threeparttable}
\end{table*}

\section{FALCON Execution Flow}\label{overview}
Consider such a scenario that a doctor wants to learn the potential disease a patient might have from an X-Ray image, only knowing the top label without the corresponding probability may lead to unreliable diagnostic result. For example, the output top label ``Pneumonia'' with probability ``0.9'' and ``0.1'' definitely have different meanings for treatment. In this section, we will outline the execution flow of FALCON.
We consider a typical convolutional neural network as shown in Fig.\ref{fig-CNN}, which contains common neural network layers. The basic idea of  FALCON is to use homomorphic encryption and Yao's Garbled Circuits to allow private neural network evaluation between two parties.

\smallskip\noindent\textbf{System model.}
We consider a client $\textit{C}$ who wants to predict an input (e.g. an X-Ray image) with a convolutional neural network model held by a server $\textit{S}$. For client $\textit{C}$, the input is private. For server $\textit{S}$, the parameters of convolutional and fully connected layers are also private. Our design goal is to preserve privacy for both parties when evaluating CNN models.
We assume that both $\textit{C}$ and $\textit{S}$ are $semi-honest$. That is, they adhere to the execution flow defined by FALCON protocols, while trying to learn the other party's private information as much as possible.

\smallskip\noindent\textbf{Privacy guarantees.}
For server $\textit{S}$, FALCON protects the following information about the model: all the weight parameters of convolutional and fully-connected layers, and the filter size of convolutional layers. FALCON does not hide the model architecture, i.e., the type of layer, layer size (the number of neurons in a layer), and the number of layers. For client $\textit{C}$, FALCON leaks no information about the input content but does not protect the input size.
We show the comparison for privacy guarantees with related works in Table \ref{table:privacy}, where all schemes use two semi-honest parties. We can see that FALCON preserves as much privacy as GAZELLE and outperforms  other works in privacy preserving.

\smallskip
At the beginning, $\textit{C}$ holds an input vector \textbf{x} and the private key, and $\textit{S}$ holds the neural network model. To evaluate the first layer, which is mostly a convolutional layer, $\textit{C}$ encrypts the FFT of \textbf{x}, denoted by $[\mathcal{F}(\textbf{x})]$, and transfers it to $\textit{S}$. Then, $\textit{S}$ and $\textit{C}$ together do the following:
\begin{enumerate}[leftmargin=17pt]
	\item \textbf{(Evaluate the Conv layer)} $\textit{S}$ feeds the convolutional layer with $[\mathcal{F}(\textbf{x})]$ and obtains the output $[\mathcal{F}(\textbf{y})]$, where \textbf{y} is the plaintext output. In order to compute the next activation layer, $\textit{S}$ and $\textit{C}$ will each hold an additive share of \textbf{y}. To this end, $\textit{S}$ uses a random vector \textbf{r} to mask the ciphertext to obtain $[\mathcal{F}(\textbf{y-r})]$ and sends it to $\textit{C}$. The client $\textit{C}$ decrypts it and performs the inverse FFT to obtain \textbf{y-r}. The client $\textit{C}$ sets its share $\textbf{x}^\textit{C}=\textbf{y-r}$ and the server $\textit{S}$ sets its share $\textbf{x}^\textit{S}=\textbf{r}$. At this time, $\textit{S}$ and $\textit{C}$ additively share the output value \textbf{y}, which is also the input to the next layer. (See details in Section \ref{secure_conv}.)
	
	\item \textbf{(Evaluate the ReLU layer)} For the ReLU layer, $\textit{S}$ and $\textit{C}$ can run the boolean circuits designed for the ReLU function using Yao's Garbled Circuits. Note that, we still require that the output value is additively shared by two parties. (See  details in Section \ref{secure_relu}.)
	
	\item \textbf{(Evaluate the pooling layer)} Evaluating the mean pooling function on two additive shares is simple. We can have these two parties perform mean pooling on their shares respectively. For max pooling, we also design boolean circuits to realize it. Note that, the same as ReLU, we need to ensure that $\textit{S}$ and $\textit{C}$ additively share the output value. (See details in Section \ref{secure_pooling}.)
	
	\item \textbf{(Evaluate the FC layer)}  Typically, the fully connected layer operation is treated as the matrix multiplication. In our design, we convert this layer into a convolutional layer while preserving the correctness of output values. Then we can use the same method in the convolutional layer to evaluate this layer. Note that the input to this layer is additively shared by $\textit{S}$ and $\textit{C}$, so we need to translate from shares to ciphertexts. To this end, letting the shares of $\textit{S}$ and $\textit{C}$ be respectively $\textbf{x}^\textit{S}$ and $\textbf{x}^\textit{C}$, $\textit{C}$ encrypts the FFT of its share and sends it to $\textit{S}$. Then $\textit{S}$ adds the FFT of its share to obtain $[\mathcal{F}(\textbf{x}^\textit{C}+\textbf{x}^\textit{S})]$, which is the input form of a convolutional layer.   (See details in Section \ref{secure_fc}.)
	
	\item \textbf{(Evaluate the softmax layer)} The input to softmax function is generally the output of a fully-connected layer. In our design, we first have $\textit{S}$ and $\textit{C}$ additively share the input. Then we disassemble the softmax function into a max and an inner product function to enable the client $\textit{C}$ to obtain the target class with probability efficiently. (See details in Section \ref{secure_softmax}.)
\end{enumerate}

\section{FALCON Design}
In this section, we present privacy-preserving protocols for each layer according to the order. 

\subsection{Setup}
Before moving on to the implementation details of each layer, we first introduce the encryption method and the translation between a ciphertext and additive shares. At the beginning, the client $\textit{C}$ holds the input \textbf{x} and needs to transfer its ciphertext to the server $\textit{S}$. In our design, all ciphertexts correspond to plaintext data in the frequency domain. That is, for input \textbf{x} of size $w\times h$, the client $\textit{C}$ first performs the FFT according to Eq. \ref{fft} to obtain $\mathcal{F}(\textbf{x})$, and then encrypts it. The $\mathcal{F}(\textbf{x})$ inherits the size of \textbf{x}, but every element of it is a complex number, e.g. $\mathcal{F}(\textbf{x})_{0,0}=a+bj$ where $a$ and $b$ are real numbers. Note that we cannot apply homomorphic encryption directly to complex numbers. Thus, we let the client $\textit{C}$ encrypt the real parts (e.g., $a$) and the imaginary parts (e.g., $b$) into two ciphertexts respectively. That is, for every element in $\mathcal{F}(\textbf{x})$, $\textit{C}$ packs all the real parts into a plaintext vector and encrypts this vector, which is denoted as $[\mathcal{F}(\textbf{x})_{R}]$. Accordingly, the ciphertext of all the imaginary parts is denoted as $[\mathcal{F}(\textbf{x})_{I}]$. All the ciphertexts involved in the FALCON have this form.

The output of a linear layer, i.e., convolutional and fully connected layer, is a ciphertext, while the input to a non-linear layer is additive shares. Therefore, before feeding the output of a linear layer into a non-linear layer, we need to translate from a ciphertext to additive shares. Assume that the output of a linear layer is $[\mathcal{F}(\textbf{y})]$, which is actually $[\mathcal{F}(\textbf{y})_R]$ and $[\mathcal{F}(\textbf{y})_I]$, the aim of server $\textit{S}$ and client $\textit{C}$ is to respectively obtain $\textbf{x}^\textit{S}$
and $\textbf{x}^\textit{C}$, satisfying $\textbf{x}^\textit{S}+\textbf{x}^\textit{C}=\textbf{y}$, and guaranteeing no information about \textbf{y} will be exposed to either $\textit{S}$ or $\textit{C}$. In order to achieve this goal, $\textit{S}$ generates a random vector \textbf{r} of the same size to \textbf{y}, and performs the FFT to obtain $\mathcal{F}(\textbf{r})_R$ and $\mathcal{F}(\textbf{r})_I$. Using the SIMDAdd, $\textit{S}$ adds these values to the ciphertext homomorphically to obtain $[\mathcal{F}(\textbf{y})_R - \mathcal{F}(\textbf{r})_R]$ and $[\mathcal{F}(\textbf{y})_I - \mathcal{F}(\textbf{r})_I]$. Recall the linearity of the FFT shown in Eq. \ref{fft_linearity}, we have
$$[\mathcal{F}(\textbf{y})_R - \mathcal{F}(\textbf{r})_R] = [\mathcal{F}(\textbf{y}-\textbf{r})_R]$$ and
$$[\mathcal{F}(\textbf{y})_I - \mathcal{F}(\textbf{r})_I] = [\mathcal{F}(\textbf{y}-\textbf{r})_I].$$
The client $\textit{C}$ decrypts them and combines the imaginary parts with the real parts to obtain $\mathcal{F}(\textbf{y}-\textbf{r})$, and then performs the inverse FFT to get $(\textbf{y}-\textbf{r})$.
Letting $\textbf{x}^\textit{S}=\textbf{r}$ and $\textbf{x}^\textit{C}=(\textbf{y}-\textbf{r})$, we have the \textbf{y} be additively shared.

To translate from additive shares to a ciphertext, namely feed the output of a non-linear layer to a linear layer, we can run the reverse of the above process. Assume that $\textit{C}$ holds $(\textbf{y}-\textbf{r})$ and $\textit{S}$ holds \textbf{r}, the client $\textit{C}$ encrypts the FFT of its share and sends the ciphertext to $\textit{S}$; the server $\textit{S}$ homomorphically adds the FFT of its share to obtain $[\mathcal{F}(\textbf{y})]$.

Note that FALCON works in $\mathbb{Z}_p$, where $p$ is the selected plaintext module for lattice-based homomorphic encryption, while the original neural networks need to perform floating-point calculations on decimal numbers. We use a simple solution which has been used in previous works to handle this issue, i.e. treating decimal numbers as integers by proper scaling and dropping the fractional parts. This way of dealing with decimal numbers only leads to a negligible loss of accuracy, which will be shown in Section \ref{sec:performance}. 
For any intermediate value $x_m$, $x_m<\lfloor{p/2}\rfloor$ implies $x_m$ is positive, otherwise negative.

\subsection{Secure Convolutional Layer}\label{secure_conv}
Recall that the input to a convolutional layer is the ciphertext of \textbf{x} of size $w \times h \times c$. In our setting, there are two ciphertexts, namely $[\mathcal{F}(\textbf{x})_R]$ and $[\mathcal{F}(\textbf{x})_I]$, and both have size $w \times h \times c$. Also, the selected $n$, i.e. the number of plaintext slots, satisfied $n > w\times h$. The server $\textit{S}$ has access to $k$ plaintext filters and needs to compute the convolution results of the input and these filters. Without FFT, previous works transform the convolution operations to matrix multiplications first, and then calculate it on the ciphertext domain. But the computational complexity of this approach is too high, because the number of additions and multiplications it takes grows with the size of the input and filters. Even if uses SIMD fashion, it still involves many heavy permutation operations in the ciphertext domain \cite{juvekar2018gazelle}. In FALCON, by utilizing the FFT, we reduce the computational complexity, where the server $\textit{S}$ only needs to perform SIMD homomorphic additions (SIMDAdd) and scalar multiplications (SIMDMul). In particular, the number of SIMDAdd and SIMDMul is respectively fixed to 2 and 4 for a given input image.

At a high level, client $\textit{C}$ and  server $\textit{S}$ transform their raw data, i.e. the input and the filters, into the frequency domain with the FFT, and then $\textit{S}$ evaluates the convolutional layer in the frequency domain with homomorphic encryption. In this part, we first introduce a simple case where the input has single channel ($c=1$) and the layer has only one filter ($k=1$) to present our key idea. Then we describe a more general case where $c>1$ and $k>1$.

\begin{figure*}[htp]
	\centering
	\includegraphics[width=0.9\textwidth]{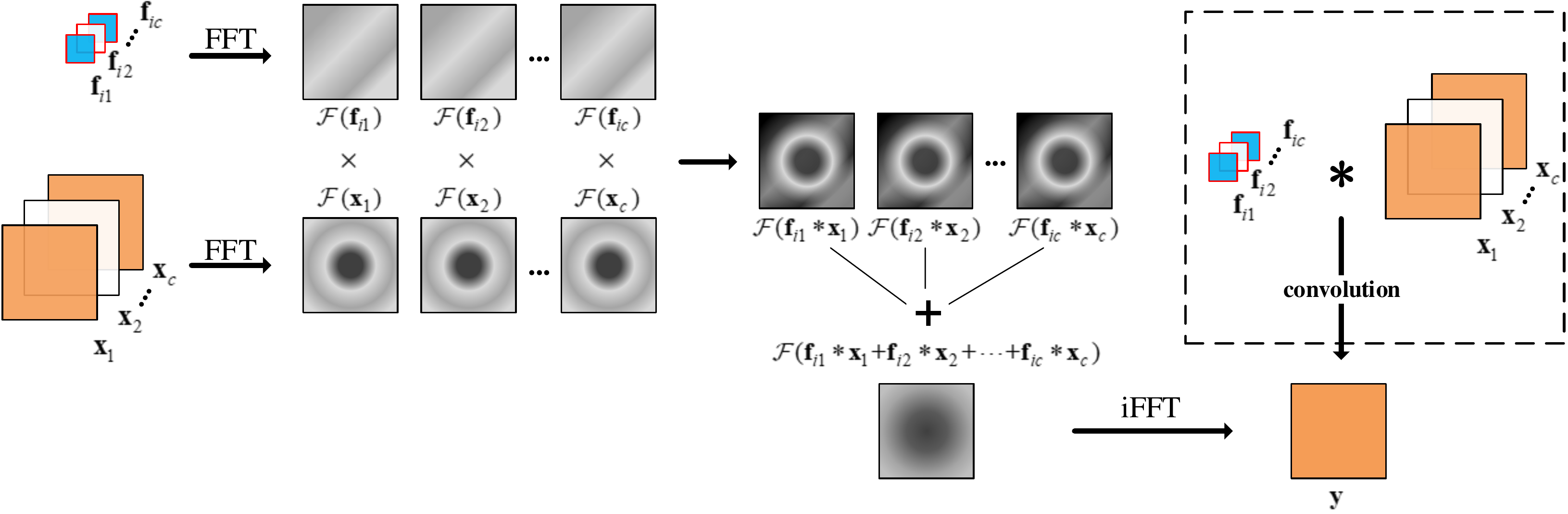}
	\caption{The convolution operations for multiple channels in plaintext.}
	\label{fig-conv_genral}
\end{figure*}

\subsubsection{Simple Case.}
In this case, the inputs are two ciphertexts, namely $[\mathcal{F}(\textbf{x})_R]$ and $[\mathcal{F}(\textbf{x})_I]$, both plaintexts of which have size $w \times h$. To apply the convolution operation on frequency domain,  server $\textit{S}$ performs the FFT on the filter of size $f_w \times f_h$ to obtain the result of size $w \times h$. Recall that the output of the FFT has the same shape with the input, and thus $\textit{S}$ first pads the filters with zeros until its size reaches $w \times h$, and then performs the FFT. Letting $\textbf{f}_i$ denote the filter, we can get the FFT results of the filter as $\mathcal{F}(\textbf{f}_i)_R$ and $\mathcal{F}(\textbf{f}_i)_I$. Note that, the number of plaintext slots is large enough to encode $w \times h$ elements into one plaintext vector, i.e. $w \times h <n$. According to the \textit{Convolution Theorem} (\ref{conv_theorem}), the pointwise product of $\mathcal{F}(\textbf{x})$ and $\mathcal{F}(\textbf{f}_i)$ corresponds to the convolution of $\textbf{x}$ and $\textbf{f}_i$. Assume that the convolution result is \textbf{y} and the corresponding frequency domain values are $\mathcal{F}(\textbf{y})_R$ and $\mathcal{F}(\textbf{y})_I$, we have
$$
\begin{aligned}
\left[\mathcal{F}(\textbf{y})_R\right] = [\mathcal{F}(\textbf{x})_R] \otimes \mathcal{F}(\textbf{f}_i)_R\hspace{4pt} &\oplus\hspace{4pt} [\mathcal{F}(\textbf{x})_I] \otimes (-\mathcal{F}(\textbf{f}_i)_I),\\
[\mathcal{F}(\textbf{y})_I] = [\mathcal{F}(\textbf{x})_R] \otimes \mathcal{F}(\textbf{f}_i)_I\hspace{4pt}&\oplus\hspace{4pt} [\mathcal{F}(\textbf{x})_I] \otimes \mathcal{F}(\textbf{f}_i)_R,
\end{aligned}
$$
where ``$\otimes$'' represents SIMDMul, and ``$\oplus$'' represents SIMDAdd. The server $\textit{S}$ then generates a random vector \textbf{r} of size $w \times h$, and encrypts its FFT values as $[-\mathcal{F}(\textbf{r})_R]$ and $[-\mathcal{F}(\textbf{r})_I]$. Finally, $\textit{S}$ sends the following two ciphertexts to $\textit{C}$:
$$
\begin{aligned}
\left[\mathcal{F}(\textbf{y}-\textbf{r})_R\right] &= \left[\mathcal{F}(\textbf{y})_R\right] \oplus [-\mathcal{F}(\textbf{r})_R], \\
\left[\mathcal{F}(\textbf{y}-\textbf{r})_I\right] &= \left[\mathcal{F}(\textbf{y})_I\right] \oplus [-\mathcal{F}(\textbf{r})_I].
\end{aligned}
$$

The client $\textit{C}$ decrypts the ciphertexts, combines the real parts with the imaginary parts, and performs the inverse FFT to obtain $(\textbf{y}-\textbf{r})$, which is set to $\textit{C}$'s share $\textbf{x}^\textit{C}$. Then, server $\textit{S}$ sets \textbf{r} to its share $\textbf{x}^\textit{S}$. At this point, the convolutional layer has been evaluated, and the result \textbf{y} is additively shared by $\textit{S}$ and $\textit{C}$.

\subsubsection{General Case.}
In the convolutional layer, all $k$ filters are independently convolved with the input to get $k$ outputs, as shown in Fig. \ref{fig-conv}. Thus, we take the convolution implementation for a filter of size $f_w \times f_h \times c$ and an input of size $w\times h\times c$ as an example. In order to present our idea clearly, we first explain how to calculate the convolution in the plaintext domain for the filter $\textbf{f}_i$ ($i\in\left[1, k\right]$) that contains $c$ channels $\textbf{f}_{i1},\textbf{f}_{i2},\cdots,\textbf{f}_{ic}$, and the input \textbf{x} that contains $c$ channels $\textbf{x}_{1},\textbf{x}_{2},\cdots,\textbf{x}_{c}$. As shown in Fig. \ref{fig-conv_genral}, $c$ $\times$ 2-D filters and $c$ $\times$ 2-D inputs are first transformed using the FFT, and then the corresponding channels are multiplied to get $c$ intermediate results. Finally, these intermediate results are added to obtain the final output in the frequency domain.

According to the plaintext calculation process, we can think of a straightforward calculation process for the ciphertext. That is, consider these 2-D filters and inputs as $c$ independent groups, and apply the calculation process in the above simple case to them to get $c$ intermediate ciphertexts. Adding these ciphertexts together, we can obtain the final ciphertext result. However,  this method wastes many plaintext slots since $w \times h < n$, and sometimes even $w \times h << n$. Also,  client $\textit{C}$ has to transfer $c$ ciphertexts to  server $\textit{S}$, which takes up more bandwidth.

In order to make full use of the plaintext slots, a heuristic approach is to pack as much data as possible into a plaintext vector, and thus there will be $2 \cdot \lceil (w \times h \times c) / n \rceil$ ciphertexts, where ``2'' is due to the fact that the real parts and the imaginary parts are packed separately. A follow-up question is that we need to add the intermediate results inside ciphertexts due to our packing. This can be done by using homomorphic permutation, which has a high computation overhead. Fortunately, in our design, we let  server $\textit{S}$ return the intermediate ciphertexts after masking with random vector \textbf{r} to  client $\textit{C}$ , who decrypts these ciphertexts and adds them in the plaintext domain. After that, $\textit{S}$ and $\textit{C}$ set their $\textbf{x}^\textit{S}$ and $\textbf{x}^\textit{C}$ respectively to additively share the output \textbf{y}.

\subsubsection{Security analysis.}
The input data of client $\textit{C}$, weight parameters, and the size of filters require protection in FALCON. Since the input data remain encrypted during the evaluation of server $\textit{S}$, the data are protected. Client $\textit{C}$ only obtains the masked convolutional result,  and thus learns nothing about the weight parameters. Because filters are padded into the same size with the input, their size is also preserved.

\subsection{Secure Activation Layer \& Pooling Layer}\label{secure_act_pool}
We next describe the implementation for activation and pooling layers together, and we will introduce a method to accelerate these two layers by adjusting the processing pipeline to reduce the number of required operations. To be noted, although using secure two-party computation to securely evaluate non-linear layers has been proposed and used in previous works, e.g., GAZELLE, there lacks implementation details. We implement the boolean circuits for non-linear layers with one of the most advanced secure two-party computation library, ABY framework. 
In what follows, we first introduce the data preprocessing phase, which translates the additive shares of $\textit{C}$ and $\textit{S}$ to Yao sharing, and is used in ABY framework \cite{demmler2015aby} for Yao's Garbled Circuits. The data preprocessing also guarantees that the Yao sharing ranges from $\left[0,p\right)$. Then we present the boolean circuits for the ReLU and max pooling functions. This boolean circuits ever was used in GAZELLE and we present it here for the completeness of the paper. Also, since the ciphertext is decrypted by the client which may reveal information about the server's input, we should also guarantee the circuit privacy. This has been addressed in GAZELLE and we thus omit the details here. The optimized solution for these two layers is given in what follows. Finally, we describe the method to translate Yao sharing back to additive shares.
Before introduing the technical details, we declare that the data preprocessing only happens when shifting from convolutional layers to activation or pooling layers. That is, for consecutive activation or pooling layers, the data prepocessing will be performed only once. 

\subsubsection{Data preprocessing.}
Assume that $\textbf{x}^\textit{C}=\{x^\textit{C}_1,x^\textit{C}_2,\cdots, x^\textit{C}_N\}$ and $\textbf{x}^\textit{S}=\{x^\textit{S}_1,x^\textit{S}_2,\cdots,x^\textit{S}_N\}$ are the additive shares held by  client $\textit{C}$ and  server $\textit{S}$, respectively. Since both $x^\textit{C}_i$ and $x^\textit{S}_i$ belong to $\left[0,p\right)$, we have $x^\textit{C}_i + x^\textit{S}_i$ belongs to $\left[0,2p\right)$. Because our plaintext module is set to $p$, we need to limit the sum of the two input shares to the range of $0$ to $p$. This is achieved by using ABY framework to implement the boolean circuits, as shown in Listing \ref{lst-Data_Preprocessing}.

The first \textit{ADDGate} receives two inputs $x^\textit{C}$ and $x^\textit{S}$ and outputs the sum of them. The second \textit{GTGate} judges if the sum exceeds $p$, and the result is sent to the \textit{MUXGate}, which is a multiplexer that decides whether to subtract $p$ or not. The subtraction is done by the \textit{SUBGate}. These circuits guarantee that the output is $\left(x^\textit{C} + x^\textit{S}\right) \mbox{mod}~p$.  Note that all the operations on the shares are in SIMD programming style, which is supported by the ABY framework, thereby reducing the overhead greatly. In Listing \ref{lst-Data_Preprocessing}, all the \textbf{yshr} variables are vectors of length $N$, which equals to the number of input elements.

\smallskip
\noindent\textbf{Performance.} As reported in ABY paper \cite{demmler2015aby}, the computation and the communication costs of \textit{ADDGate}, \textit{SUBGate}, \textit{GTGate} and \textit{MUXGate} are almost the same when using Yao sharing. Thus, we denote the cost of one operation by \textit{SIMD(N)} , where \textit{N} is the number of values in one share, and the total cost of data preprocessing is $4\cdot\textit{SIMD(N)}$.

\begin{figure}[tp]
	\centering
	\includegraphics[width=0.3\textwidth]{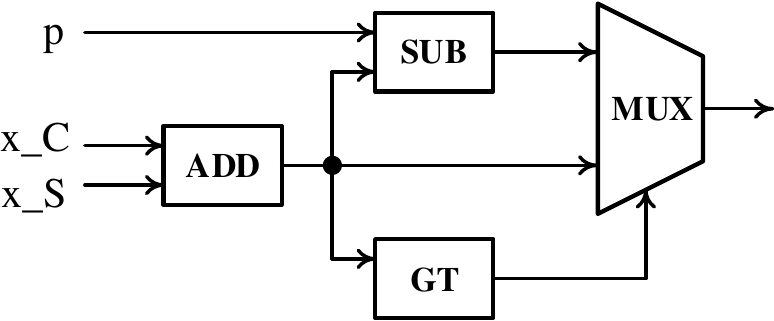}
	\caption{Boolean circuits for data preprocessing.}
	\label{fig-data_preprocessing}
\end{figure}

\begin{lstlisting}[label={lst-Data_Preprocessing},caption={Function description of data preprocessing.},captionpos=b, float=tp]
/*data preprocessing, returns (x_C+x_S)%p in Yao Sharing*/
yshr DataPreprocessing(yshr x_C, yshr x_S, yshr p){
	yshr x, gt, dif, res;
	
	x = ADDGate(x_C, x_S);
	gt = GTGate(x, p);
	dif = SUBGate(x, p);
	res = MUXGate(x, dif, gt);
	
	return res;
}
\end{lstlisting}

\begin{lstlisting}[label={lst-ReLU},caption={Function description of ReLU.},captionpos=b, float=th]
/*
ReLU, returns max(x, 0);
x is the output of DataPreprocessing();
p_2 is p/2.
*/
yshr ReLU(yshr x, yshr p_2){
yshr gt, res;
	
	gt = GTGate(x, p_2);
	res = MUXGate(x, 0, gt);
	
	return res;
}
\end{lstlisting}

\subsubsection{Secure ReLU layer.}\label{secure_relu}
Typically, a convolutional layer and a non-last fully connected layer are followed by a ReLU activation layer, which decides whether an input value should be continued to pass down, or in other words, whether the neuron is activated by the input value, as shown in Fig.~\ref{fig-opt_maxpooling_relu}. Recall that the ReLU function is $f(\textbf{x})=\mbox{max}(\textbf{x}, 0)$, and in our setting, the input $\textbf{x}=\{x_1,x_2,\cdots,x_N\}$ is additively shared by client $\textit{C}$ and server $\textit{S}$, i.e. $\textbf{x}^\textit{C}+\textbf{x}^\textit{S}=\textbf{x}$. Our aim is to enable that $\textit{C}$ and $\textit{S}$ additively share $\mbox{max}(\textbf{x}, 0)$. That is, $\textit{C}$ holds $\mbox{max}(\textbf{x}, 0)-\textbf{r}$ while $\textit{S}$ holds \textbf{r}, where \textbf{r} is randomly generated by $\textit{S}$.
MiniONN considers $\mbox{max}(\textbf{x}, 0) = \textbf{x}\cdot \mbox{compare}(\textbf{x}, 0)$, and implements the ReLU function by GTGate and MULGate (Multiplication Gate) provided by ABY. Due to the relatively high overhead of MULGate, the MULGate is replaced by MUXGate, which has a much less overhead as stated in ABY paper. The pseudocode is shown in Listing \ref{lst-ReLU}.

Recall that, for a given $x$, $x > p/2$ implies $x$ is negative, and positive otherwise. The first \textit{GTGate} performs a great-than operation ($>$), and the output is used by the \textit{MUXGate} to select the positive $x$ or $0$ as the result.

\smallskip
\noindent\textbf{Performance.} The total cost is $2\cdot\textit{SIMD(N)}$.

\subsubsection{Secure pooling layer.}\label{secure_pooling}
\begin{figure}[tp]
	\centering
	\includegraphics[width=0.46\textwidth]{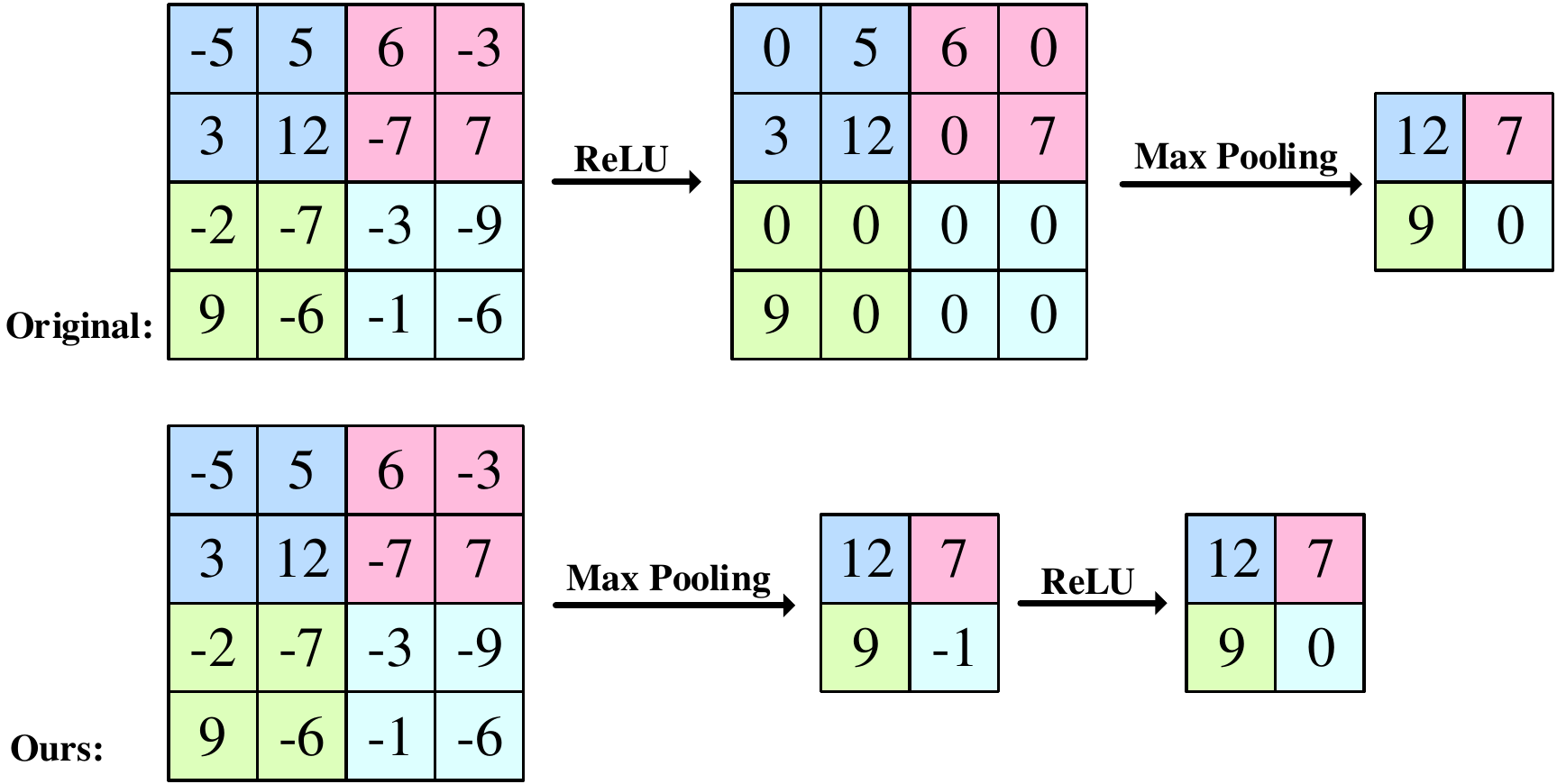}
	\caption{Original ReLU and Max Pooling v.s. Our Max Pooling and ReLU.}
	\label{fig-opt_maxpooling_relu}
\end{figure}

The pooling layer performs down-sampling by dividing the input into rectangular pooling regions and computing the mean or maximum of each region, which reduces the spatial dimensions, as shown in Fig. \ref{fig-opt_maxpooling_relu}. To evaluate the mean pooling, we can simply let client $\textit{C}$ and  server $\textit{S}$ compute the mean value of their respective shares of each region. Our focus is the evaluation of max pooling function which is implemented in Yao's Garbled Circuits provided by the ABY framework. Assume that the rectangle pooling region has size $k$, and the number of regions is approximate $N/k$, which is also the output size. For example, $N=16$, $k=2\times 2=4$ and $N/k=4$ in Fig. \ref{fig-opt_maxpooling_relu}.
Letting $\textbf{x}_{\mbox{region}}=\{x_1,x_2,\cdots,x_k\}$ be one of the rectangular pooling regions, our aim is to calculate $\max(x_1,x_2,\cdots,x_k)$. The pseudocode of the designed boolean circuits is shown in Listing \ref{lst-MaxPooling}.

Since the input has been limited from $0$ to $p/2$ by the ReLU layer, we can iteratively compare two elements to obtain the max element with \textit{GT} and \textit{MUX} circuits without considering the existence of negative elements. Because comparisons are performed inside each region, we pack $N$ elements into $k$ vectors of size $N/k$ through \textit{SubsetGate}.

\smallskip
\noindent\textbf{Performance. } The total cost is $1\cdot\textit{SubsetGate}+(k-1)\cdot\textit{SIMD(N/k)}$.

\begin{lstlisting}[label={lst-MaxPooling},caption={Function description of max pooling.},captionpos=b, float=t]
/*
MaxPooling, returns max(x1,x2,...,xk);
x is the output of ReLU();
p_2 is p/2; k is the region size.
*/
yshr MaxPooling(yshr x, yshr p_2, int k){
	yshr x_reg[k], gt, res;
	
	// split the x into k pieces.
	x_reg = SubsetGate(x);
	
	res = x_reg[0];
	for (int i = 1; i < k; i++) {
		gt = GTGate(res, x_reg[i]);
		res = MUXGate(res, reg[i], gt);
	}
	return res;
}
\end{lstlisting}

\begin{lstlisting}[label={lst-OptMaxPoolingReLU},caption={Function description of our max pooling and ReLU.},captionpos=b,numbers=left, float=t]
/*
Optimized MaxPooling and ReLU;
x is the output of DataPreprocessing();
p_2 is p/2; k is the region size.
*/
yshr OptMaxPoolingAndReLU(yshr x, yshr p, yshr p_2, int k){
	yshr x_reg[k], gt, res;
	
	// Firstly, perform max pooling.
	x_reg = SubsetGate(x);
	res = x_reg[0];
	for (int i = 1; i < k; i++) {
		gt = GTGate(res, x_reg[i]);
		res = MUXGate(res, reg[i], gt);
	}
	
	//Secondly, perform ReLU.
	gt = GTGate(res, p_2);
	res = MUXGate(res, 0, gt);
	
	return res;
}
\end{lstlisting}

\subsubsection{The Optimized ReLU and max pooling layers.}
Recall that in a typical processing pipeline, a ReLU layer is followed by a max pooling layer, and the basic operation of both is $\max()$. Assume that $\textbf{x}_{\mbox{region}}=\{x_1,x_2,\cdots,x_k\}$ is one of the rectangle pooling regions but has not applied to the ReLU function. Then, the final output of the max pooling layer should be
$$
\max\left(\max\left(x_1,0\right),\max\left(x_2,0\right),\cdots,\max\left(x_k,0\right)\right),
$$
where the inside $\max()$ corresponds to the ReLU function while the outside is the max pooling function.We can find that this process is equivalent to the following one:
$$
\max\left(\max\left(x_1,x_2,\cdots,x_k\right),0\right),
$$
where the inside $\max()$ can be considered as the max pooling function while the outside as the ReLU function. Based on this observation, reversing the position of the ReLU layer and the max pooling layer in the processing pipeline will reduce the number of $\max()$ operations. An example is shown in Fig.~\ref{fig-opt_maxpooling_relu}. In fact, this trick has been proposed in the study of deep learning \cite{executionorder}. Nevertheless, due to the fact that the ReLU and max pooling functions are relatively much cheaper than the heavy convolutional and fully connected layers in the plaintext domain, this optimization has been discarded. However, in the ciphertext domain, all these functions have great impacts on the overall performance. We report and utilize this optimization here to further improve the FALCON performance.
With this approach, we can see that the number of $\max()$ operations in the max pooling layer does not change, but the ReLU layer is reduced greatly. For a max pooling layer with $(2\times 2)$ region with a stride of 2, our method can save 75\% of the ReLU operations.

The pseudocode is shown in Listing \ref{lst-OptMaxPoolingReLU}. We first apply max pooling operations to obtain max values of each region, and ReLU operations follow to filter out all negative values.

\smallskip
\noindent\textbf{Performance.} The total cost is $1\cdot\textit{SubsetGate}+(k-1)\cdot\textit{SIMD(N/k)}+2\cdot\textit{SIMD(N/k)}=1\cdot\textit{SubsetGate}+(k+1)\cdot\textit{SIMD(N/k)}$. If we use the original implementation of ReLU and Max pooling operations, the cost would be $2\cdot\textit{SIMD(N)}+1\cdot\textit{SubsetGate}+(k-1)\cdot\textit{SIMD(N/k)}$. Thus, our acceleration approach can reduce by $2\cdot\textit{SIMD(N)}-2\cdot\textit{SIMD(N/k)}$.

\subsubsection{Output circuits for ReLU and Max Pooling}
The original outputs of the ReLU or the max pooling circuits are in the form of Yao sharing. Our aim is to additively share the result between client $\textit{C}$ and  server $\textit{S}$. This can be achieved by taking the output \textbf{y} and a random vector $(-\textbf{r}~\mbox{mod}~p)$ generated by $\textit{S}$ as two inputs of the data preprocessing circuit, and the result of which is $(\textbf{y}-\textbf{r}~\mbox{mod}~p)$ and is sent to $\textit{C}$.

\subsubsection{Security analysis.}
Since ReLU and max pooling layers do not have private model parameters, we only focus on the confidentiality of the input. Due to the security of Yao's Garbled Circuits, the input data are hidden.

\subsection{Secure Fully Connected Layer}\label{secure_fc}
Normally, the fully connected layer can be treated as the multiplication of a weight matrix and an input vector, and this can be executed very fast in the plaintext domain. However, in the ciphertext domain, especially using SIMD techniques in homomorphic encryption, this kind of multiplication is expensive. Inspired by the observation that fully connected layer can be viewed as convolutional layers with filters that cover the entire input regions \cite{long2015fully}, we propose an efficient solution by transforming the FC layer to the convolutional layer first, and then utilizing the acceleration method in Section \ref{secure_conv} to evaluate the fully-connected layer.

We here briefly describe how to transform a fully connected layer into a convolutional layer. Recall that a fully connected layer first flattens the input of size $w \times h \times c$ to a vector of size $l_i$, where $l_i=w\times h\times c$, and then multiplies it with a matrix of size $l_0\times l_i$ to obtain the output vector of size $l_0$. For such a fully connected layer, we can transform it to a convolutional layer with $l_0$ filters, each of which has size $w\times h\times c$, i.e. $f_w=w$ and $f_h=h$, and remove the flatten operation. The size of the output will be $1\times 1\times l_0$, which can be viewed as a vector of length $l_0$.

The next steps are straightforward. We can use the proposed secure convolution method introduced in Section \ref{secure_conv} to evaluate the transformed fully connected layer. Note that, for classification CNNs, the last layer is always a fully connected layer, the output of which is treated as a weight vector, also known as \textit{logits}. Each value in this vector represents the weight of the corresponding label in the classification result. Therefore, $\textit{C}$ can run an oblivious \textit{argmax} using Yao's Gabled Circuits with $\textit{S}$ to obtain the result.

\medskip
\noindent\textbf{Security analysis.}
The fully-connected layer is evaluated exactly as a convolutional layer, and thus the weight parameters are also protected and the the client cannot tell the difference between a covolutional and a fully connected layer.

\subsection{Secure Softmax Layer}\label{secure_softmax}
In all the classification CNNs, the last fully connected layer is always followed by a softmax layer to generate the probability distribution over $K$ different possible classes. However, to our best knowledge, in all previous work, the authors presented that the server can return the \textit{logits} to the client, who could obtain the probabilities by performing softmax function locally, e.g. GAZELLE, or the client runs \textit{argmax} using secure two-party computation to only obtain the classification result without knowing \textit{logits} and probabilities, e.g. MiniONN. There are two main reasons why these schemes bypass the encrypted computation of softmax layer:
\begin{enumerate}[leftmargin=17pt]
	\item Implementing the softmax function would bring very high computation complexity, no matter using homomorphic encryption or secure two-party protocols.
	\item The papers did not realize that revealing the \textit{logits} or probabilities of all classes may reveal a lot of data of the model.
\end{enumerate}
The heavy computation complexity is due to the division and exponentiation operations in the softmax function when using homomorphic encryption or secure two-party computation, and we thus propose a division and exponentiation free protocol to avoid performing these operations with secure two-party computation. We notice that in a client-server scenario, by only accessing prediction results, $\textit{C}$ is able to extract an equivalent or near-equivalent model \cite{tramer2016stealing}, infer the training set \cite{shokri2017membership, salem2018ml}, and generate adversarial examples \cite{carlini2017towards}. To tackle this issue, $\textit{S}$ can only return the necessary result, i.e. the class to which the input belongs  and its corresponding probability, to $\textit{C}$. In MiniONN, the target class will be returned without probability.
While in FALCON, the probability can also be obtained by $\textit{C}$ with our secure softmax protocol.

Recall that the softmax function is given by
$$
f(\textbf{x})_i=\frac{e^{x_i}}{\sum_{k=1}^{K}{e^{x_k}}}, \quad\mbox{for $i=1,2,\cdots,K$},
$$
where $f(\textbf{x})_i$ is the probability that the input belongs to the class $i$.
Letting the target class be $t$, our aim is to calculate
$
p_t=\frac{e^{x_t}}{\sum_{k=1}^{K}{e^{x_k}}}.
$
For the numerator, we have the client $\textit{C}$ calculate \textit{argmax} and \textit{max} functions with the server $\textit{S}$ using Yao's Garbled Circuits to learn $t$ and share $x_t$. For the denominator, we let $\textit{S}$ and $\textit{C}$ multiplicatively share $e^{\textbf{x}}$, thus translating highly complex calculations into simple vector inner product. Furthermore, we ignore some items in $e^{\textbf{x}}$ to improve the performance while guaranteeing the accuracy of $p_t$, and we will prove its correctness.

Before moving to the detailed protocols, we first give the following theorem:
\begin{theorem}\label{theorem1}
	For $p_t=\frac{e^{x_t}}{\sum_{k=1}^{K}{e^{x_k}}}$, where $x_t=\max(x_1,\cdots,\\x_K)$, and $p_{t}'=\frac{e^{x_t}}{\sum\limits_{x_k \geq x_t - m}{e^{x_k}}}$, where $m\geq ln\left[(10^l-1)(K-1)\right]$ and $l\geq 1$, we have $\left|p_t-p_t'\right|\leq 10^{-l}$.
\end{theorem}
The proof of correctness can be found in the appendix \ref{proof}. This theorem shows that in the case of an accuracy requirement of $10^{-l}$, we can replace $p_t$ with $p_t'$. In another word, we can set a threshold $x_t-m$, and all the values less than the threshold will be ignored during the calculation, which can significantly reduce the computational complexity. Furthermore, we can transform $p_t'$ as follows

$$
\begin{aligned}
p_{t}'&=\frac{e^{x_t}}{\sum\limits_{x_k \geq x_t - m}{e^{x_k}}}=\frac{e^{x_t}/e^{x_t-m}}{\sum\limits_{x_k \geq x_t - m}{e^{x_k}}/e^{x_t-m}}\\
&=\frac{e^{m}}{\sum\limits_{x_k \geq x_t - m}{e^{x_k-(x_t-m)}}},
\end{aligned}
$$
where all the intermediate exponential values are limited to $\left[e^0,e^m\right]$, which enables us to use a small bit length to evaluate the secure softmax with Yao's Garbled Circuits. For example, for an accuracy requirement of $10^{-3}$ and the number of classes $K$ is 100, we have $m\geq ln(10^{-3}*100-100)\approx 11.52$, and $e^{12}$ takes only 18 bits, while the original $x_t$ may reach up to $>$ 100 \cite{carlini2017towards} and $e^{100}$ takes 145 bits.

Based on the above analysis, the outline protocol of our proposed secure softmax is as follows:
\begin{enumerate}[leftmargin=17pt]
	\item  Let  $[\mathcal{F}(\textbf{x})]$, where $\textbf{x}=\{x_1,x_2,\cdots,x_K\}$, be the input to the softmax layer and the output of the fully connected layer. Server $\textit{S}$ masks it with a random vector \textbf{r}, and sends $[\mathcal{F}(\textbf{x}-\textbf{r})]$ to client $\textit{C}$. Then $\textit{S}$ sets its share to $\textbf{x}^\textit{S}=\textbf{r}=\{{r_1},{r_2},\cdots,{r_K}\} ~\mbox{mod}~p$.
	\item The client $\textit{C}$ decrypts $[\mathcal{F}(\textbf{x}-\textbf{r})]$ and performs the inverse FFT to obtain $(\textbf{x}-\textbf{r})$. Then $\textit{C}$ set its share to $\textbf{x}^\textit{C}={\textbf{x}-\textbf{r}}=\{{x_1-r_1},{x_2-r_2},\cdots,{x_K-r_K}\} ~\mbox{mod}~p$.
	\item Now $\textit{C}$ and $\textit{S}$ interact with each other to find the maximum value $x_t$ and decide which $x_i$ can be ignored according to the selected integer $m$, and set the ignored one and the left $x_i$ to $0$ and $m-(x_t-x_i)$, respectively. To be noted, the plaintext modulo is converted to $(m+1)$, and there no longer exist negative values. At the end of this procedure, $\textit{C}$ and $\textit{S}$ hold newly generated shares, $\textbf{x}^\textit{S}=\{{r_1'},{r_2'},\cdots,{r_K'}\} ~\mbox{mod}~(m+1)$ and $\textbf{x}^\textit{C}=\{{x_1'-r_1'},{x_2'-r_2'},\cdots,{x_K'-r_K'}\} ~\mbox{mod}~m+1$, where $x_i'$ is $0$ or $m-(x_t-x_i)$ and $r_i'$ is randomly generated by $\textit{S}$.
	\item Next, to calculate the denominator of $p_t'$, client $\textit{C}$ and  server $\textit{S}$ first calculate $e^{\textbf{x}^\textit{C}}=\{e^{x_1'-r_1'},e^{x_2'-r_2'},\cdots,e^{x_K'-r_K'}\}$ and
	$e^{\textbf{x}^\textit{S}}=\{e^{r_1'},e^{r_2'},\cdots,e^{r_K'}\}$. Then, they use Yao's Garbled Circuits to calculate the denominator of $p_t'$. The boolean circuits used here for Yao's Garbled Circuits can be simply implemented with \textit{ADDGate}, \textit{MULGate} and \textit{MUXGate}, and we ignore the details here.
	The $\textbf{x}^\textit{C}$ and  $\textbf{x}^\textit{S}$ are used to guarantee that every $e^{(x_i'-r_i'~\mbox{mod}~m+1) + (r_i'~\mbox{mod}~m+1)}$ does not exceed $e^{m+1}$ and decide whether to drop it. The final calculation result, i.e. the denominator of $p_t'$, will be obtained by $\textit{C}$.
\end{enumerate}

At this point, since $\textit{C}$ owns the denominator of $p_t$ and the numerator $e^m$ is public, it is able to calculate $p_t'$.

\medskip
\noindent{\textbf{Security analysis.}}
Since all the private information, i.e. $x_1,x_2,\cdots,x_K$, remains shared throughout the protocol, their privacy is guaranteed. Note that, the potential leakage resulting from $p_t'$ is out of scope of this paper. Actually, it is a general problem of neural networks \cite{sharif2016accessorize}. Since the numerator of $p_t'$, i.e. $e^m$, is public and contains no information about $p_t'$, the possible information that could infer from the denominator of $p_t'$ is equal to $p_t'$.

\section{Performance Evaluation}\label{sec:performance}
We implemented FALCON in C++. For fast Fourier Transform (FFT), we used FFTW library \cite{fftw}. For additively homomorphic encryption, we used the Fan-Vercauteren (FV) scheme \cite{bajard2016full}.
For secure two-party computing, we used Yao's Garbled Circuits implemented by the ABY framework. For performance comparison purpose, we used the same parameters $p$ and $q$ as what used in \cite{juvekar2018gazelle}. The selected parameters for both schemes are 128-bit security level. Specifically, the number of slots $n$, the plaintext module $p$, and the ciphertext module $q$ are needed for the initialization of the FV scheme. We chose $n=2048$, which means we can process up to 2048 elements in parallel, and the ciphertext module $q$ was set to 1152921504382476289. The plaintext module $p$ was set to 1316638721, which has 30-bit length and is enough for all the intermediate values since they can be securely scaled down as discussed in Section \ref{overview} and suggested by MiniONN\cite{liu2017oblivious}. 

We tested FALCON on two computers, both of which are equipped with Intel i5-7500 CPU with 4 3.40 GHz cores and 8GB memory, and have Ubuntu 16.04 installed. We let one be the client $\textit{C}$ and the other play as the server $\textit{S}$. We took experiments in the LAN setting similar to previous work \cite{liu2017oblivious, juvekar2018gazelle}. Each experiment was repeated for 100 times and we report the mean in this paper.
In this section, we will first evaluate the benchmarks for individual layers in the order mentioned in our contributions, and then compare FALCON with previous work on real-world models. As the authors in GAZELLE and MiniONN haven't released their codes, here we use the results from their papers for comparison.


\subsection{Benchmarks for Layers}\label{benchmark_layers}
Here we introduce the performance of different layers, i.e. convolutional, fully-connected, ReLU, max pooling, and softmax layers. Since GAZELLE is the best known one among existing works, we only compare FALCON with it in all layers except the softmax layer, which is not implemented by GAZELLE. For benchmarking, all input data to each layers are randomly sampled from $\left[0, p\right)$. For convolutional and fully connected layers, their parameters are chosen from the CIFAR-10 model stated in Section \ref{evalutation-realworld}.

Our first contribution is accelerating convolutional and fully-connected layers by the integrated use of FFT and lattice-based homomorphic encryption. Thus we present the benchmarks for convolutional and fully-connected layers in the different input size. As shown in Table~\ref{table:conv}, for convolutional layers, we show the online running time with input ($w\times h\times c$) and filter ($f_w\times f_h\times c$, $k$) using different frameworks. For fully-connected layers, we report the running time with the input vector of length $l_i$ and the output vector of length $l_o$. Note that, the setup phases involve performing FFT on filters and encrypting random values for masking, and the online phases take only the server's computation into account. As one can see from Table~\ref{table:conv}, FALCON outperforms GAZELLE in both setup and online phases. Especially for online phases, our efficient Conv and FC implementations offer us over $10\times$ less runtime.

\begin{table}[h]
	\caption{Benchmarks and Comparisons for Conv and FC.}
	\label{table:conv}
	\centering
	\resizebox{\linewidth}{!}{
		\begin{tabular}{c|ccc|c|c}
			\hline
			\multirow{2}{*}{\textbf{Layer}} & \multirow{2}{*}{\textbf{Input}}& \multirow{2}{*}{\textbf{Filter/Output}} & \multirow{2}{*}{\textbf{Framework}} & \multicolumn{2}{|c}{\textbf{Time} (ms)} \\
			\cline{5-6}
			&&&&setup &online \\
			\hline\hline
			\multirow{3}{*}{\tabincell{c}{Conv\\Layer}} & ($28\times 28\times 1$) & ($5\times 5\times 1$, 5) & \tabincell{c}{GAZELLE\\FALCON} & \tabincell{c}{11.4\\3.1} & \tabincell{c}{9.2\\0.25}\\
			\cline{2-6}
			& ($16\times 16\times 128$) & ($3\times 3\times 128$, 128) & \tabincell{c}{GAZELLE\\FALCON} & \tabincell{c}{3312\\615} & \tabincell{c}{704\\51.2}\\
			\hline\hline
			\multirow{3}{*}{\tabincell{c}{FC\\Layer}} & 2048 & 1 & \tabincell{c}{GAZELLE\\FALCON} & \tabincell{c}{16.2\\1.2} & \tabincell{c}{8.0\\0.1}\\
			\cline{2-6}
			& 1024 & 16& \tabincell{c}{GAZELLE\\FALCON} & \tabincell{c}{21.8\\9.6} & \tabincell{c}{7.8\\0.8}\\
			\hline
	\end{tabular}}
\end{table}

In Table \ref{table:relu}, we report the running time and communication overhead of setup and online phases for ReLU and Max Pooling layers. Note that, since we utilize ABY to implement our scheme, these experimental results succeed its optimization for secure two-party computation.  In addition to the data preprocessing, the online communication overhead of ReLU and max pooling operations is almost negligible. We can also see that the optimized max pooling and ReLU operations have reduced the computation and communication overhead in all phases. Therefore, in ReLU and MaxPooling layers, FALCON which uses the optimized version outperforms GAZELLE which uses the original version.

\begin{table}[h]
	\caption{Benchmarks for ReLU and Max Pooling.}
	\label{table:relu}
	\centering
	\resizebox{\linewidth}{!}{
		\begin{tabular}{c|c|c|c|c|c}
			\hline
			\multirow{2}{*}{\textbf{Operation}} & \multirow{2}{*}{\tabincell{c}{\textbf{Number of} \\\textbf{Inputs}}} & \multicolumn{2}{|c|}{\textbf{Time} (ms)}&  \multicolumn{2}{|c}{\textbf{Comm} (MB)}\\
			\cline{3-6}
			&&setup &online &setup& online\\
			\hline\hline
			\tabincell{c}{Data\\Preprocessing} & \tabincell{c}{1000 \\ 10000} & \tabincell{c}{32.3 \\ 265.5} & \tabincell{c}{14.5 \\ 136.4} & \tabincell{c}{4.82 \\ 48.2} & \tabincell{c}{1.45\\ 14.9}\\
			\cline{2-5}
			\hline
			\multirow{1}{*}{ReLU} & \tabincell{c}{1000 \\ 10000} & \tabincell{c}{9.82 \\ 96.2} & \tabincell{c}{4.20 \\ 43.2} & \tabincell{c}{1.95 \\ 19.2} & \tabincell{c}{0.01\\ 0.11}\\
			\cline{2-5}
			\hline
			\multirow{1}{*}{MaxPooling} & \tabincell{c}{1000 \\ 10000} & \tabincell{c}{12.1 \\ 100} & \tabincell{c}{5.6 \\ 45.5} & \tabincell{c}{1.94 \\ 20.0} & \tabincell{c}{0.01\\ 0.12}\\
			\cline{2-5}
			\hline\hline
			\multirow{1}{*}{ReLU+MaxPooling} & \tabincell{c}{1000 \\ 10000} & \tabincell{c}{21.9 \\ 196.3} & \tabincell{c}{9.8 \\ 88.7} & \tabincell{c}{3.89 \\ 39.2} & \tabincell{c}{0.02\\ 0.23}\\
			\cline{2-5}
			\hline
			\tabincell{c}{Optimized \\MaxPooling+ReLU} & \tabincell{c}{1000 \\ 10000} & \tabincell{c}{12.5 \\ 134.3} & \tabincell{c}{5.2 \\ 54.4} & \tabincell{c}{2.44 \\ 24.4} & \tabincell{c}{0.02\\ 0.14}\\
			\cline{2-5}
			\hline
	\end{tabular}}
\end{table}

We tested the performance of our proposed protocol for softmax function in different settings. As shown in Table~\ref{table:softmax}, the runtime cost and communication overhead of setup and online phases grow with the accuracy and the number of classes. The runtime and communication overhead of our proposed softmax evaluation method is relatively small compared with other layers of a CNN.

\begin{table}[h]
	\caption{Benchmarks for the Softmax.}
	\label{table:softmax}
	\centering
	\begin{tabular}{c|c|c|c|c|c}
		\hline
		\multirow{2}{*}{\textbf{Accuracy}} & \multirow{2}{*}{\textbf{Classes}} & \multicolumn{2}{|c|}{\textbf{Time} (ms)}&  \multicolumn{2}{|c}{\textbf{Comm} (MB)}\\
		\cline{3-6}
		&&setup &online &setup& online\\
		\hline\hline
		$10^{-2}$ & \tabincell{c}{10\\ 100 \\ 1000} & \tabincell{c}{8.56\\58.7\\574.8} & \tabincell{c}{3.89\\24.5\\254.6} & \tabincell{c}{0.996\\9.96\\99.6} & \tabincell{c}{0.0294\\0.294\\29.4}\\
		\cline{2-5}
		\hline
		$10^{-4}$ & \tabincell{c}{10\\ 100 \\ 1000} & \tabincell{c}{8.66\\60.3\\588.0} & \tabincell{c}{4.02\\26.7\\257.6} & \tabincell{c}{0.996\\9.96\\99.6} & \tabincell{c}{0.0294\\0.294\\29.4}\\
		\cline{2-5}
		\hline		
	\end{tabular}
\end{table}

\subsection{Evaluations on Real-World Models}\label{evalutation-realworld}
We evaluated the performance of FALCON on two real-world models, MNIST and CIFAR-10. We show the time cost and the communication overhead in both setup and online phases. The time cost includes the computation cost of client $\textit{C}$ and server $\textit{S}$ and the transmission latency between them. To be noted, for the fairness of comparison, the softmax layer is excluded in both models.

MNIST is a dataset for handwritten digits and is widely used for image classification and recognition tasks. We use the CNN model for MNIST from \cite{liu2017oblivious}. This CNN model takes a gray scale image with size $28\times 28$ as input, and outputs the classification result, which is either 0 to 9. It has 2 convolutional, 2 fully-connected, 3 ReLU and 2 max pooling layers.

CIFAR-10 is a small RGB image dataset in 10 classes and is widely used for image classification tasks. We use the CNN model for CIFAR-10 from \cite{liu2017oblivious}. The input to this CIFAR-10 model is a three channel image of size $32\times 32\times 3$, and the output is the class it belongs to. This model contains 7 convolutional, 1 fully-connected, 7 ReLU and 2 mean pooling layers.

\begin{table}[h]
	\caption{Performance Comparison on MNIST and CIFAR-10.}
	\label{table:mnist}
	\centering
	\resizebox{\linewidth}{!}{
		\begin{tabular}{c|c|c|c|c|c|c|c}
			\hline
			\multirow{2}{*}{\textbf{CNN}} & \multirow{2}{*}{\textbf{Framework}} & \multicolumn{3}{|c|}{\textbf{Time} (s)}&  \multicolumn{3}{|c}{\textbf{Comm} (MB)}\\
			\cline{3-8}
			&&setup &online & total &setup& online & total\\
			\hline\hline
			MNIST& \tabincell{c}{MiniONN \\GAZELLE\\ FALCON} & \tabincell{c}{3.58\\0.48 \\ 0.63} & \tabincell{c}{5.74 \\0.33\\ 0.21} & \tabincell{c}{9.32 \\ 0.81 \\ 0.84} & \tabincell{c}{20.9 \\47.5\\ 77.0} & \tabincell{c}{636.6\\22.5 \\ 15.5} & \tabincell{c}{657.5\\70.0 \\ 92.5}\\
			\hline
			CIFAR-10& \tabincell{c}{MiniONN\\GAZELLE \\ FALCON} & \tabincell{c}{472 \\9.34\\ 7.20} & \tabincell{c}{72 \\3.56\\ 2.88} & \tabincell{c}{544\\12.9 \\ 10.1} & \tabincell{c}{3046 \\940\\ 1194} & \tabincell{c}{6226 \\296\\ 265} & \tabincell{c}{9272\\1236 \\ 1459}\\
			\hline
	\end{tabular}}
\end{table}

As shown in Table \ref{table:mnist}, when evaluating the online overhead on MNIST model,  FALCON is running over $27\times$ faster than MiniONN while reducing communication overhead by over 97\%. For the online overhead in CIFAR-10, FALCON is about $25\times$ faster than MiniONN and reduce communication overhead by over 97\%. The significant improvement in running time is due to the repeatedly use of FFT and lattice-based homomorphic encryption, which saves many multiplications over ciphertexts.

When comparing with GAZELLE, FALCON also shows improvements on online overhead. Meanwhile, FALCON can output the probability of the top label while GAZELLE cannot. Here we would like to make a comment about GAZELLE. We found two versions of GAZELLE, i.e., \cite{juvekar2018gazelle} and \cite{juvekar2018gazellearxiv}, and some experimental results in these two versions are not consistent. The online running time of ReLU and MaxPooling in \cite{juvekar2018gazelle} is about $10\times$ faster than that in \cite{juvekar2018gazellearxiv} while all algorithms and experimental settings stay no change. Furthermore, we also found inconsistent data  in \cite{juvekar2018gazelle}, the most recent version. In Table 11 of \cite{juvekar2018gazelle}, the offline running time of GAZELLE on network topology D is 0.481 seconds. Note that the ReLU layer of neural network D has $\approx$ 10,000 outputs and the MaxPooling layer has $\approx$ 2560 outputs, which will cost over 0.97 seconds according to Table 10 in \cite{juvekar2018gazelle} (much larger than the stated 0.481 seconds). We tried to communicate with the authors to check such inconsistency in their paper, and we did not get a reply from the authors. A comparison with GAZELLE will be added once their benchmark is finalized or they release the sourcecode. Anyway, since FALCON has better performance in each layer of CNN layers, we are confident that FALCON should outperform GAZELLE on a complete CNN.

\subsection{Prediction Accuracy on Real-World Models}
Since we treat decimal numbers as integers by proper scaling, there might have accuracy concerns on the encrypted models. Fig.~\ref{table:accuracy} shows that the loss of accuracy in FALCON is negligible, and different schemes achieve nearly the same results. Since GAZELLE did not report their models' accuracy, we only perform the comparison with MiniONN.

\begin{table}[h]
	\caption{Prediction Accuracy on MNIST and CIFAR-10.}
	\label{table:accuracy}
	\centering
	\resizebox{\linewidth}{!}{
		\begin{tabular}{c|c|c|c|c}
			\hline
			& Plaintext & MiniONN & GAZELLE & FALCON \\
			\hline\hline
			MNIST & 99.31\% &99.0\% & - & 99.26\% \\
			\hline
			CIFAR-10 & 81.61\% & 81.61 \% & - & 81.61\% \\
			\hline
	\end{tabular}}
\end{table}

\section{Conclusion}
In this paper, we presented a fast and secure evaluation approach for convolutional neural networks. For linear layers including convolutional and fully-connected, our fast Fourier Transform based scheme achieves a low latency performance. For non-linear layers including ReLU and max pooling, we provided a detailed implementation including the usage of optimized processing pipeline. For the softmax layer that has not been studied in previous works, we introduced the first efficient and privacy-preserving protocol for it. Finally, our evaluation results show that FALCON has less computation and communication overhead in each layer and achieves best known performance thus far on CNN models.

{\normalsize \bibliographystyle{acm}
	\bibliography{bibliography}}

\begin{thebibliography}{10}

\bibitem{executionorder}
Execution order of relu and max-pooling.
\newblock \url{https://github.com/tensorflow/tensorflow/issues/3180}.
\newblock Accessed Nov. 12, 2018.

\bibitem{abadi2016deep}
{\sc Abadi, M., Chu, A., Goodfellow, I., McMahan, H.~B., Mironov, I., Talwar,
  K., and Zhang, L.}
\newblock Deep learning with differential privacy.
\newblock In {\em Proceedings of the 23th ACM Conference on Computer and
  Communications Security (CCS'16)\/} (2016), ACM, pp.~308--318.

\bibitem{abdel2014convolutional}
{\sc Abdel-Hamid, O., Mohamed, A.-r., Jiang, H., Deng, L., Penn, G., and Yu,
  D.}
\newblock Convolutional neural networks for speech recognition.
\newblock {\em IEEE/ACM Transactions on Audio, Speech, and Language Processing
  22}, 10 (2014), 1533--1545.

\bibitem{amodei2016deep}
{\sc Amodei, D., Ananthanarayanan, S., Anubhai, R., Bai, J., Battenberg, E.,
  Case, C., Casper, J., Catanzaro, B., Cheng, Q., Chen, G., et~al.}
\newblock Deep speech 2: End-to-end speech recognition in english and mandarin.
\newblock In {\em Proceedings of the 33rd International Conference on Machine
  Learning (ICML'16)\/} (2016), pp.~173--182.

\bibitem{bajard2016full}
{\sc Bajard, J.-C., Eynard, J., Hasan, M.~A., and Zucca, V.}
\newblock A full rns variant of fv like somewhat homomorphic encryption
  schemes.
\newblock In {\em Proceedings of the 23rd International Conference on Selected
  Areas in Cryptography (SAC'16)\/} (2016), Springer, pp.~423--442.

\bibitem{bogdanov2008sharemind}
{\sc Bogdanov, D., Laur, S., and Willemson, J.}
\newblock Sharemind: A framework for fast privacy-preserving computations.
\newblock In {\em Proceedings of the 13th European Symposium on Research in
  Computer Security (ESORICS'08)\/} (2008), Springer, pp.~192--206.

\bibitem{bos2013improved}
{\sc Bos, J.~W., Lauter, K., Loftus, J., and Naehrig, M.}
\newblock Improved security for a ring-based fully homomorphic encryption
  scheme.
\newblock In {\em Proceedings of the 14th IMA International Conference on
  Cryptography and Coding (IMACC'13)\/} (2013), Springer, pp.~45--64.

\bibitem{brakerski2014leveled}
{\sc Brakerski, Z., Gentry, C., and Vaikuntanathan, V.}
\newblock (leveled) fully homomorphic encryption without bootstrapping.
\newblock {\em ACM Transactions on Computation Theory 6}, 3 (2014),
  13:1--13:36.

\bibitem{carlini2017towards}
{\sc Carlini, N., and Wagner, D.}
\newblock Towards evaluating the robustness of neural networks.
\newblock In {\em Proceedings of the 38th IEEE Symposium on Security and
  Privacy (SP'17)\/} (2017), pp.~39--57.

\bibitem{googlevision}
{\sc Cloud, G.}
\newblock Vision api - image content analysis.
\newblock \url{https://cloud.google.com/vision/}, 2018.
\newblock [Accessed 4th-May-2018].

\bibitem{demmler2015aby}
{\sc Demmler, D., Schneider, T., and Zohner, M.}
\newblock Aby-a framework for efficient mixed-protocol secure two-party
  computation.
\newblock In {\em Proceedings of the 22nd Annual Network and Distributed System
  Security Symposium(NDSS'15)\/} (2015).

\bibitem{esteva2017dermatologist}
{\sc Esteva, A., Kuprel, B., Novoa, R.~A., Ko, J., Swetter, S.~M., Blau, H.~M.,
  and Thrun, S.}
\newblock Dermatologist-level classification of skin cancer with deep neural
  networks.
\newblock {\em Nature 542}, 7639 (2017), 115--118.

\bibitem{fan2012somewhat}
{\sc Fan, J., and Vercauteren, F.}
\newblock Somewhat practical fully homomorphic encryption.
\newblock {\em IACR Cryptology ePrint Archive 2012\/} (2012), 144.

\bibitem{fftw}
{\sc FFTW}.
\newblock Fast fourier transform.
\newblock \url{http://www.fftw.org}, 2018.
\newblock [Accessed 4th-May-2018].

\bibitem{gentry2012fully}
{\sc Gentry, C., Halevi, S., and Smart, N.~P.}
\newblock Fully homomorphic encryption with polylog overhead.
\newblock In {\em Proceedings of the 31st Annual International Conference on
  the Theory and Applications of Cryptographic Techniques (EUROCRYPT'12)\/}
  (2012), Springer, pp.~465--482.

\bibitem{gilad2016cryptonets}
{\sc Gilad-Bachrach, R., Dowlin, N., Laine, K., Lauter, K., Naehrig, M., and
  Wernsing, J.}
\newblock Cryptonets: Applying neural networks to encrypted data with high
  throughput and accuracy.
\newblock In {\em Proceedings of the 33rd International Conference on Machine
  Learning (ICML'16)\/} (2016), pp.~201--210.

\bibitem{goldreich1987play}
{\sc Goldreich, O., Micali, S., and Wigderson, A.}
\newblock How to play any mental game.
\newblock In {\em Proceedings of the 19th Annual ACM Symposium on Theory of
  Computing (STOC'87)\/} (1987), ACM, pp.~218--229.

\bibitem{hadji2018we}
{\sc Hadji, I., and Wildes, R.~P.}
\newblock What do we understand about convolutional networks?
\newblock {\em arXiv preprint arXiv:1803.08834\/} (2018).

\bibitem{he2016deep}
{\sc He, K., Zhang, X., Ren, S., and Sun, J.}
\newblock Deep residual learning for image recognition.
\newblock In {\em Proceedings of the 29th IEEE Conference on Computer Vision
  and Pattern Recognition (CVPR'16)\/} (2016), pp.~770--778.

\bibitem{juvekar2018gazellearxiv}
{\sc Juvekar, C., Vaikuntanathan, V., and Chandrakasan, A.}
\newblock Gazelle: A low latency framework for secure neural network inference.
\newblock Available at \url{https://arxiv.org/pdf/1801.05507.pdf} (uploaded by
  Chiraag Juvekar on 16 Jan 2018).

\bibitem{juvekar2018gazelle}
{\sc Juvekar, C., Vaikuntanathan, V., and Chandrakasan, A.}
\newblock {GAZELLE}: A low latency framework for secure neural network
  inference.
\newblock In {\em Proceedings of the 27th {USENIX} Security Symposium ({USENIX}
  Security'18)\/} (2018).
\newblock Available at
  \url{https://www.usenix.org/conference/usenixsecurity18/presentation/juvekar}.

\bibitem{krizhevsky2012imagenet}
{\sc Krizhevsky, A., Sutskever, I., and Hinton, G.~E.}
\newblock Imagenet classification with deep convolutional neural networks.
\newblock In {\em Advances in Neural Information Processing Systems
  (NIPS'12)\/} (2012), pp.~1097--1105.

\bibitem{liu2017oblivious}
{\sc Liu, J., Juuti, M., Lu, Y., and Asokan, N.}
\newblock Oblivious neural network predictions via {MiniONN} transformations.
\newblock In {\em Proceedings of the 24th ACM Conference on Computer and
  Communications Security (CCS'17)\/} (2017), ACM, pp.~619--631.

\bibitem{long2015fully}
{\sc Long, J., Shelhamer, E., and Darrell, T.}
\newblock Fully convolutional networks for semantic segmentation.
\newblock In {\em Proceedings of the IEEE conference on computer vision and
  pattern recognition(CVPR'15)\/} (2015), pp.~3431--3440.

\bibitem{mohassel2017secureml}
{\sc Mohassel, P., and Zhang, Y.}
\newblock Secureml: A system for scalable privacy-preserving machine learning.
\newblock In {\em Proceedings of the 38th IEEE Symposium on Security and
  Privacy (SP'17)\/} (2017), IEEE, pp.~19--38.

\bibitem{nair2010rectified}
{\sc Nair, V., and Hinton, G.~E.}
\newblock Rectified linear units improve restricted boltzmann machines.
\newblock In {\em Proceedings of the 27th International Conference on Machine
  Learning (ICML'10)\/} (2010), pp.~807--814.

\bibitem{riazi2018chameleon}
{\sc Riazi, M.~S., Weinert, C., Tkachenko, O., Songhori, E.~M., Schneider, T.,
  and Koushanfar, F.}
\newblock Chameleon: A hybrid secure computation framework for machine learning
  applications.
\newblock {\em arXiv preprint arXiv:1801.03239\/} (2018).

\bibitem{rouhani2017deepsecure}
{\sc Rouhani, B.~D., Riazi, M.~S., and Koushanfar, F.}
\newblock Deepsecure: Scalable provably-secure deep learning.
\newblock {\em arXiv preprint arXiv:1705.08963\/} (2017).

\bibitem{salem2018ml}
{\sc Salem, A., Zhang, Y., Humbert, M., Fritz, M., and Backes, M.}
\newblock Ml-leaks: Model and data independent membership inference attacks and
  defenses on machine learning models.
\newblock {\em arXiv preprint arXiv:1806.01246\/} (2018).

\bibitem{sharif2016accessorize}
{\sc Sharif, M., Bhagavatula, S., Bauer, L., and Reiter, M.~K.}
\newblock Accessorize to a crime: Real and stealthy attacks on state-of-the-art
  face recognition.
\newblock In {\em Proceedings of the 23rd ACM Conference on Computer and
  Communications Security (CCS'16)\/} (2016), ACM, pp.~1528--1540.

\bibitem{shen2017deep}
{\sc Shen, D., Wu, G., and Suk, H.-I.}
\newblock Deep learning in medical image analysis.
\newblock {\em Annual Review of Biomedical Engineering 19\/} (2017), 221--248.

\bibitem{shokri2015privacy}
{\sc Shokri, R., and Shmatikov, V.}
\newblock Privacy-preserving deep learning.
\newblock In {\em Proceedings of the 22nd ACM Conference on Computer and
  Communications Security (CCS'15)\/} (2015), ACM, pp.~1310--1321.

\bibitem{shokri2017membership}
{\sc Shokri, R., Stronati, M., Song, C., and Shmatikov, V.}
\newblock Membership inference attacks against machine learning models.
\newblock In {\em Proceedings of the 38th IEEE Symposium on Security and
  Privacy (SP'17)\/} (2017), pp.~3--18.

\bibitem{song2017machine}
{\sc Song, C., Ristenpart, T., and Shmatikov, V.}
\newblock Machine learning models that remember too much.
\newblock In {\em Proceedings of the 24th ACM Conference on Computer and
  Communications Security (CCS'17)\/} (2017), ACM, pp.~587--601.

\bibitem{tramer2016stealing}
{\sc Tram{\`e}r, F., Zhang, F., Juels, A., Reiter, M.~K., and Ristenpart, T.}
\newblock Stealing machine learning models via prediction apis.
\newblock In {\em Proceedings of the 25th USENIX Security Symposium (USENIX
  Security'16)\/} (2016), pp.~601--618.

\bibitem{winograd1978computing}
{\sc Winograd, S.}
\newblock On computing the discrete fourier transform.
\newblock {\em Mathematics of computation 32}, 141 (1978), 175--199.

\bibitem{yao1986generate}
{\sc Yao, A. C.-C.}
\newblock How to generate and exchange secrets.
\newblock In {\em Proceedings of the 27th Annual Symposium on Foundations of
  Computer Science (FOCS'86)\/} (1986), IEEE, pp.~162--167.

\end{thebibliography}

\appendix
\section{Proof of Correctness}\label{proof}
We here provide a proof for the correctness of Theorem \ref{theorem1}. Our proof idea is to derive the range of $m$ requiring $\left|p_t-p_t'\right|\leq 10^{-l}$.
\begin{proof}
	Since $p_t=\frac{e^{x_t}}{\sum_{k=1}^{K}{e^{x_k}}} \leq \frac{e^{x_t}}{\sum\limits_{x_k \geq x_t - m}{e^{x_k}}}=p_{t}'$, we have
	$$\left|p_t-p_t'\right|\leq 10^{-l} \Leftrightarrow p_t'-p_t\leq 10^{-l}. $$
	If we require $p_t'-p_t\leq 10^{-l}$, then
	\begin{equation}\label{eq1}
	\frac{e^{x_t}}{\sum\limits_{x_k \geq x_t - m}{e^{x_k}}} - \frac{e^{x_t}}{\sum_{k=1}^{K}{e^{x_k}}} \leq 10^{-l}.
	\end{equation}
	Letting $T_0=e^{x_t}$, $T_1=\sum\limits_{\substack{x_t-x_k\leq m\\k\ne t}}{e^{x_k}}$, and $T_2=\sum\limits_{x_t-x_k>m}{e^{x_k}}$, we can rewrite inequality (\ref{eq1}) as
	\begin{equation}\label{eq2}
	\frac{T_0}{T_0+T_1} - \frac{T_0}{T_0+T_1+T_2} \leq 10^{-l},
	\end{equation}
	which is equivalent to
	\begin{equation}\label{eq3}
	\frac{T_0^2+2T_0 T_1 + T_1^2}{T_0T_2} + \frac{T_1}{T_0} + 1 \geq 10^l.
	\end{equation}
	If we want this inequality holds, then we require that the minimum of the left side is greater or equal to the right side. Since $T_0$ is fixed, we can minimize $T_1$ and maximize $T_2$ to minimize the left side of the inequality. Letting the length of $T_1$ and $T_2$ be $s_1$ and $s_2$ respectively, we have
	\begin{displaymath}
	\mbox{minimum}(T_1)=s_1\cdot e^{x_t-m},~\mbox{maximum}(T_2)=s_2\cdot e^{x_t-m}.
	\end{displaymath}
	Bringing the minimum $T_1$ and maximum $T_2$ to the inequality (\ref{eq3}), we can derive
	
	\begin{equation}\label{eq4}
	\begin{aligned}
	m \geq  ln(\sqrt{\frac{[(10^l-1)^2\cdot s_2-4\cdot 10^l\cdot s_1]\cdot s_2}{4}}\\ +\frac{(10^l-1)\cdot s_2-2\cdot s_1}{2}).
	\end{aligned}
	\end{equation}
	If we want the inequality (\ref{eq4}) holds, $m$ needs to be greater or equal to the maximum of the right side. Since $l\geq 1$, the maximum value of the right side is obtained when maximizing $s_2$ and minimizing $s_1$. Because $s_1+s_2=K-1$, we have
	\begin{displaymath}
	\mbox{minimum}(s_1)=0,~\mbox{maximum}(s_2)=K-1.
	\end{displaymath}
	Bringing the minimum $s_1$ and maximum $s_2$ to the inequality (\ref{eq4}), we have
	\begin{equation}
	m \geq  ln\left[(10^l-1)(K-1)\right].
	\end{equation}
\end{proof}

\end{document}